# MULTILINGUAL WEBSITE USABILITY ANALYSIS BASED ON AN INTERNATIONAL USER SURVEY

Mahdi H. Miraz[1,2], Maaruf Ali[2] and Peter Excell[1]

[1]Department of Computing, Glyndwr University, Wrexham, UK
`(m.miraz||p.excell)@glyndwr.ac.uk`
[2]Department of Computer Science & Software Engineering, University of Hail, KSA
`(m.miraz||m.maaruf)@uoh.edu.sa`

*ABSTRACT*

*A study was undertaken to determine the important usability factors (UF) used in the English and the non-English version of a website. The important usability factors were determined, based on a detailed questionnaire used in an international survey. Analysis of the questionnaire found inequalities in the user satisfaction and a general dissatisfaction with the non-English version of the website. The study concluded that more care should be taken in creating the text, taking into account the cultural and linguistic background of the users and the use of graphics in multilingual websites.*

## 1. INTRODUCTION

Internationalisation of services is continuing at an unabated rate. For any multinational or even nowadays any national website, it has to cater to an audience whose mother tongue is usually not English. How people interact with these websites can have a significant impact on the success and reputation of the business. Human Computer Interaction (HCI) and specifically Usability Factors play an important part in the success, usage and continued accessing of the website, success for business being mainly measured by repeated transactions.

Various Usability Factors were used in the survey to assess the user satisfaction of using a key exemplar (the BBC's news website) and its non-English counterpart. The sample population consisted of both native English speakers and bilingual and or multilingual users. The focus group consisted of people from different countries, primarily including: Bangladesh, Egypt, India, Jordan, Pakistan and Saudi Arabia.

The purpose of the study was to determine whether the Usability Factors are being optimised equally between the English and non-English versions of the websites. If any discrepancies are identified, then these are discussed with a view to identifying how to unify the user's satisfaction in the interaction with both websites.

## 2. BACKGROUND

The vast majority of websites, some 89% [1], are presently written in the English vernacular. Despite this, the share of non-English language websites is actually on the rise [2] and this strongly shows the necessity of creating non-English website content, specially with the advance of globalisation [3]. Related to making websites bilingual or multilingual is the critical aspect of ensuring both linguistic and cultural compatibility, especially so as not to cause any offence. Not only this, but the user behaviour (or the expected behaviour) and interaction has to be uniform between the English and non-English versions of the websites in order to elicit expected business and behavioural returns [4]. These issues concerning the user's satisfaction being related to his/her cultural habits have been mentioned by Hermeking [5].

The seriousness of compatibility issues in human understanding of international websites was in fact addressed by the W3C [2] organisation in October 2010. They concluded that there was a need to fill the gaps using the current technologies available to ensure the universal appeal and accessibility of websites. Miraz *et al.* [6] explored the socio-economic and cultural issues between UK and Bangladeshi web users. Earlier researchers found that different cultural users comprehended the same websites in totally different ways: misunderstanding, confusion and even offence to users could be caused by the inappropriate use of some metaphors, interaction sequences, appearance or navigation [7] [8]. These studies, though extensive, did not address the issues related to English-language websites and their multilingual versions.

Another earlier study (in 2001) [9] was limited to the multilingual English-speaking country of Botswana. Though this was a very good paper, technology has progressed much, including the level of internationalisation, globalisation and the use of the Web. The present paper thus explores the unique study of multilingual website usability based on an international user survey.

A case study approach was undertaken to study the UI Design issues as apprehended by a focus group using the BBC [10] English and non-English versions of the websites. The questionnaire was designed in bilingual format: English was the common language. The first languages of participants were also used, especially Arabic, Bangla, Hindi and Urdu. All the participants were adults aged between 18 and 80 years old. The focus group were given the task of exploring both the English and non-English (depending on the mother tongue of the users) versions of the BBC Online website. They were then required to provide their feedback concerning the design issues and usability of their website interaction, using the said questionnaire.

## 3. DESIGN ISSUES

Usability Issues that were considered covered both multilingual and multi-regional websites. The Usability Factors were derived from the usability attributes and heuristics familiarized by Neilson Jakob [11] [12], the questionnaire suggested and used by him was also adopted and modified to meet the multilingual aspect of our research. It must be stressed that this is an initial questionnaire based study that needs further development. The focus group were all professional international volunteers with no known visual perception issues. This last point is important so that the data collected is not invalid. Scaling of the websites was also considered. In this section, the User Interface (UI) design issues relating to multilingual websites will be discussed and the data collected through conducting the survey will be analyzed. The focus will be on the following issues:

- Language Selection and Usability
- Graphics and Placement of Text and Images
- Colour
- Translation
- Abbreviation and Keywords
- Localizing the website
- Page layout and Navigation
- Font Size Legibility of Websites
- Fitting the Text into Web Pages

### 3.1. Language Selection and Usability

Intelligent and automatic language localization is essential for accessing multilingual websites. The UF must take into consideration the fact that a non-native user may be residing in a country whose language s/he is not proficient in. Thus the website being delivered to him/her must have means of being able to be switched into a language that s/he can understand or prefers. One intelligent way of achieving this is by IP tracking to select the geographical region and then auto-detection of the user's browser language can take place so that the website can be automatically served in that language. A 'change back' option should always be present to facilitate the possibility that the user wishes to visit pages of some other regions or languages, or even if s/he relocates (periodically).

### 3.2. Graphics Placement and Readability Related to the Comprehension of Text and Images

People from different cultures and linguistic backgrounds read by scanning text or pictograms in different directions, either horizontally (right-to-left or left-to-right) or vertically (top-to-bottom). Thus it is very important to take all these into consideration when placing images and text in the multilingual website as they play a vital rôle for the overall usability, convenience and acceptance of a website. The survey found that over half the participants considered the graphics displayed in the English version of the website to be clear and attractive whilst the majority of users of the non-English website were unhappy, as shown in Table 1. However, the use of graphics as an aid to text comprehensibility did not show any marked difference between the English and non-English versions of the websites.

Table 1. Website's Graphics Usefulness Responses.

| Graphics clarity and attractiveness of English version of the website | English Version | | Non-English Version | |
| --- | --- | --- | --- | --- |
| | No. of Responses | Percentage | No. of Responses | Percentage |
| Yes | 84 | 51.2% | 58 | 35.4% |
| To some extent | 62 | 37.8% | 84 | 51.2% |
| No | 18 | 11.0% | 22 | 13.4% |
| Total observation | 164 | 100% | 164 | 100% |

### 3.3. Colour

Colour has always had universal symbolic significance throughout human history. However, this symbolism is not unique and the same colour may have different and often diametrically opposing significance depending on regional conventions. Thus the usage of colour in a multi-regional website should be used with particular caution. For example, testing has in fact shown that efficiency in the intelligibility and interactivity actually increased in Russia when using a black and red combination for a "Call to Action" button whilst in Italy by changing it from red to orange [13]. Our study has in fact confirmed this, with a greater user satisfaction (40%) reported in our questionnaire when using the English version colour scheme of the BBC website over its non-English versions (29%). The results are shown in Figs. 1 and 2. Our study clearly shows much thought needs to be applied when choosing a colour scheme for a multilingual website.

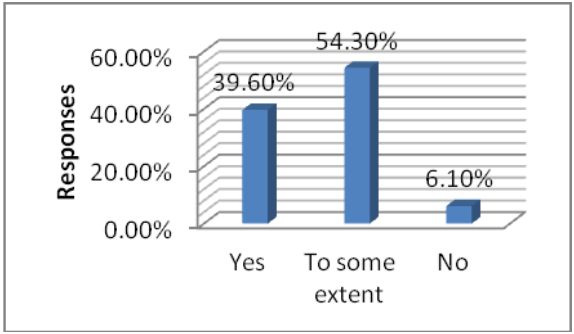 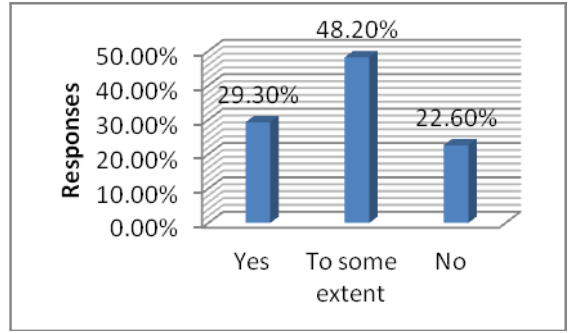

Figure 1. English Website Attractiveness and Appeal of Colours.

Figure 2. Non-English Website Attractiveness and Appeal of Colours.

### 3.4 Translation

It is a natural facet of the human psyche that personal predilections manifest themselves in a person's work. This behaviour can be seen in the work of translators, despite their training in trying to maintain neutrality. The use of automated translators unfortunately suffers from the problem of literal translation: they fail to translate the focused meaning especially when metaphors, cultural terms or terms having multiple meanings are involved.

Cap [14] conducted a study on the neutrality of Wikipedia (WP), as shown in Table 3, by analyzing how pages dated March 2010 covering a highly controversial figure were reported based on the website's geographical origin and using the example of "Bin Laden".

Table 2. Translation and Description of Bin Laden.

| Language | Reporting Perception |
|---|---|
| **German** | as a terrorist |
| **English** | a leader of a terrorist organization |
| **Arabic** | "founder and leader of al-Qaeda network" |
| **Hebrew** | "terrorist leader of the Islamic terrorist organization Al-Qaeda" |
| **Chinese** | "leader [of] the organization [...] a lot of people think that is a global terrorist organization". |

Thus deviations in the neutrality of websites can clearly be seen, based on the geographical region where the website is compiled.

Our research did in fact find a divergence in the translated versions of the English BBC website. Surprisingly, as little as 23.8% of the participants considered the translation of the Non-English version matches closely to the original news of the English version, as shown in Figure 3.

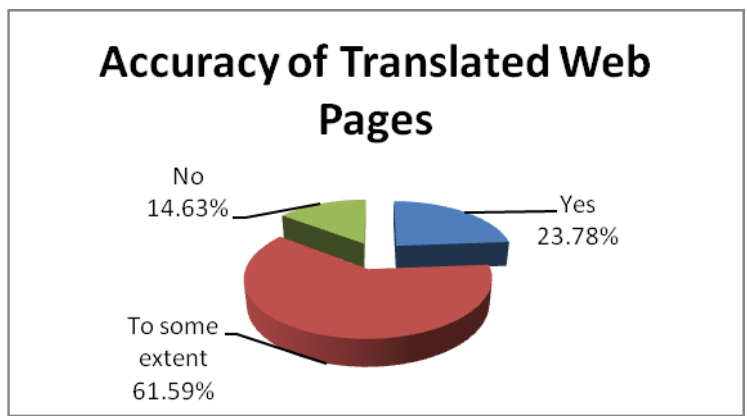

Figure 3. Perception of the Accuracy of Translated Web Pages.

### 3.5. Abbreviation and Keywords

The use of language not only extends to the use of actual words and phrases but also to the use of particular acronyms, these often are Latin in origin in the Romance Languages of the West, numbering an estimated 800 million users in the world [15]. However, this still only represents just over 11% of the world's population. Thus finding the non-English version/translation of such acronyms as *etc*., *e.g*., *i.e*., *et al*., FAQ, ASAP, are sometimes very difficult, often because they do not exist in the non-English target language. Sometimes the only solution is to use a literal translation. The survey found that only 27.4% of the participants consider the abbreviated words to have suitable alternative non-English substitutions. The full results are show in Figure 4, below.

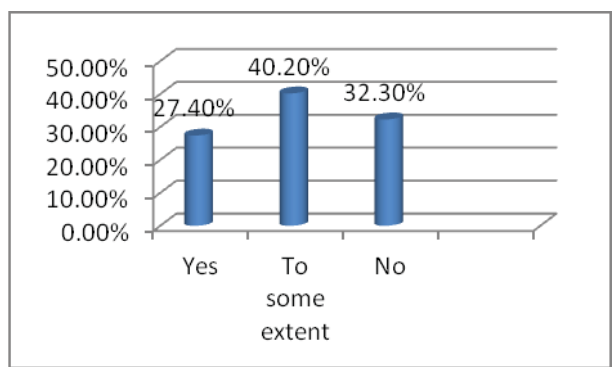

Figure 4. Existence of Abbreviations in Multiple Languages.

Similar problems are also faced when using or searching for keywords. For example "cheap flights" does not have a literal translation in Italian. Using '*voli economici*' in Italy only returns 33,000 searches per month versus 6.12 million searches per month in America for "cheap flights". However, searching for "*voli low cost*" gives a significantly higher hit rate of 246,000 searches per month [13].

Some words like "Google" do not appear to have any etymology, apart from being similar to a very large number in mathematics. Such invented words often are difficult to translate using just a single word, so a few words would be needed to explain them, unless they have become popular through widespread global usage.

### 3.6. Localizing the Website

When delivering content for a particular region, translation of the words is not the only aspect that needs to be considered. The content has to appeal to the region in question in terms of local interests and concerns such as the regional weather, regional news, local weights and measures which may not be S.I. nor imperial, currency symbols and conversions, date format, government holidays, cultural sensitivities and taboos, geographic examples and gender rôles, beliefs and religions, traditions and social structures, level of use of ICT and education, to mention a few. The survey found that 42.1% believed the localised contents that they accessed were indeed matched to their needs. However, this is still less than 50%, so there is much scope for improvement here. The results of the survey are shown in Figure 5.

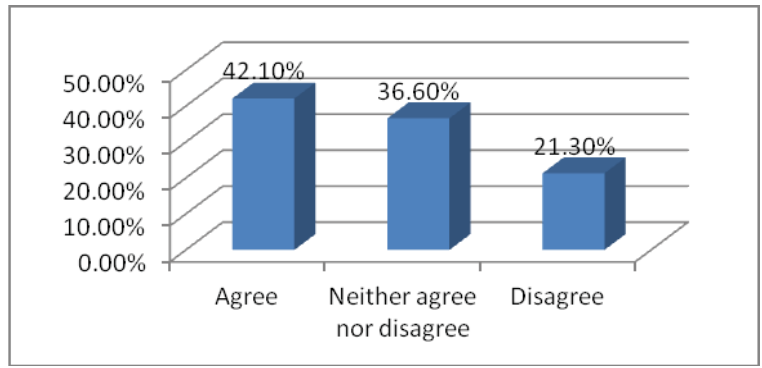

Figure 5. Language Localization of the News.

## 3.7. Page Layout and Navigation

The ability to understand the layout of a website is very important for the reader's satisfaction. Most Web users tend to have a short attention span. Based on 59,573 page views [16], most Web readers (49%) like to browse pages consisting of no more than 111 words, this equates to about 26 seconds per page (based on an adult average reading speed of 250 words per minute). This brevity of engagement emphasises the importance of trying to ensure retention of these Web readers and their repeated returns back to the same site. Our study looked at the comprehensibility of the page layout between the English and its non-English versions. The study found that there was a marginally greater dissatisfaction with the layout of the non-English websites. This is shown in Figures 6 and 7, where 8.5% of the non-English version users found the page layout difficult to comprehend, compared with only 1.2% of the English website users.

The satisfaction of the page layout can be made more uniform by adopting the same layout style but also taking into consideration the textual reading conventions of different cultures. This may sometimes just require the use of a mirror-imaged layout version of the website.

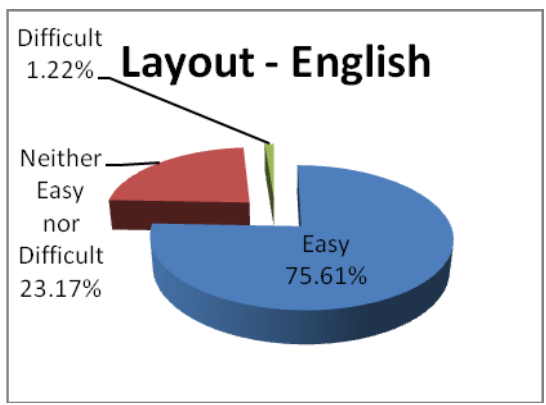
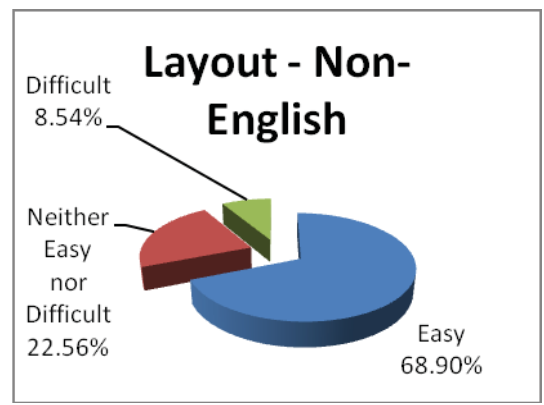

Figure 6. Web Layout – English Version.   Figure 7. Web Layout – Non-English Version.

Navigability is also a major factor when browsing a webpage. The results from Figures 8 and 9 show a marginal preference for the non-English website in terms of page navigability of 6.1% over the English version. This could be explained by the content being a somewhat streamlined version of the English website.

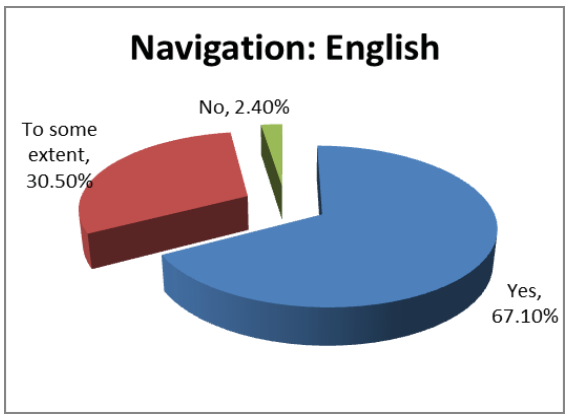
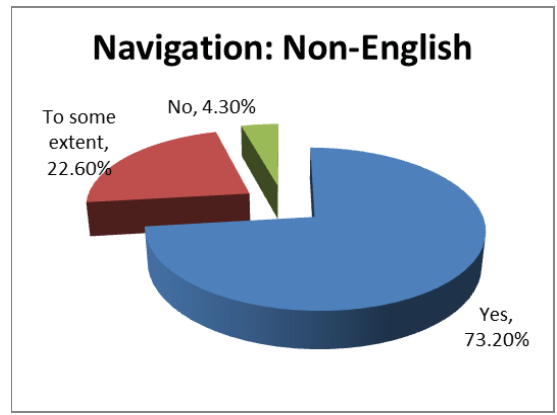

Figure 8. Navigability of English BBC Website.

Figure 9. Navigability of Non-English BBC Website.

### 3.8 Font Size Legibility of Websites

Due to the intricacies of some characters, such as Chinese and Japanese, they need to be rendered at a sufficient resolution for them to be legible. This is also the case with Arabic and other Oriental or similar based alphabets because of their use of vowel and diacritical marks. This is not usually the case with Latin or Cyrillic based characters. Thus the survey found that 67.7% of participants consider the fonts of the English version to be readable, attractive and properly sized. In contrast, 61.6% of participants considered the fonts of the non-English version to be readable. Though similar in satisfaction, there is clearly scope for improvements in the display of both languages, in particular choosing the appropriate font and size for both. The results are shown in Figures 10 and 11. Dyslexic people tend to prefer sans serif fonts with a coloured background and this usability factor needs to be seriously considered also.

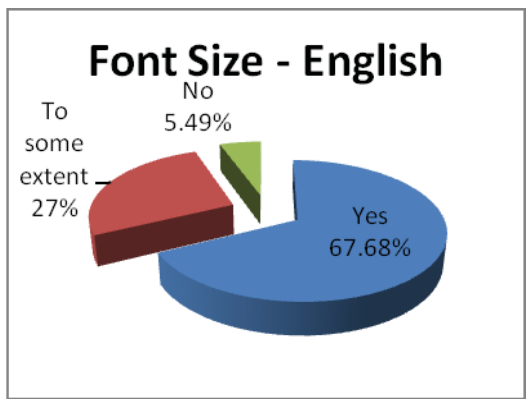
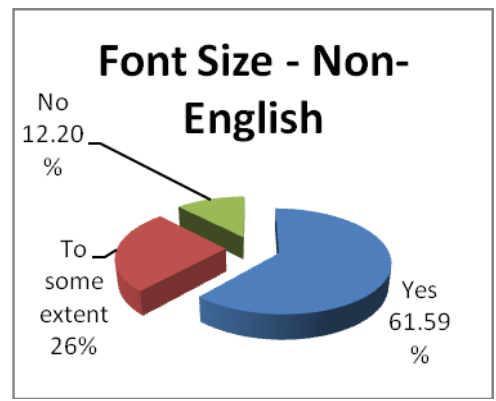

Figure 10. Font Size Legibility of English Website.

Figure 11. Font Size Legibility of Non-English Website.

### 3.9 Fitting the Text into Web Pages

Some languages can express ideas more succinctly than others and this naturally has to be carefully considered in translation and in using them in certain Web structures, especially menus that have fixed widths. For example German and Russian compared to English can take up more space, whilst Chinese and Korean take up far less space, due to their pictographic representation [17]. Our survey has found that 64.6% of the participants found no difficulty reading the English version but this percentage is lower (55.5%) in the non-English versions. One reason could be due to the resolution of the screen which implies the need for scalability to be also taken into consideration. These results are shown in Figures 12 and 13.

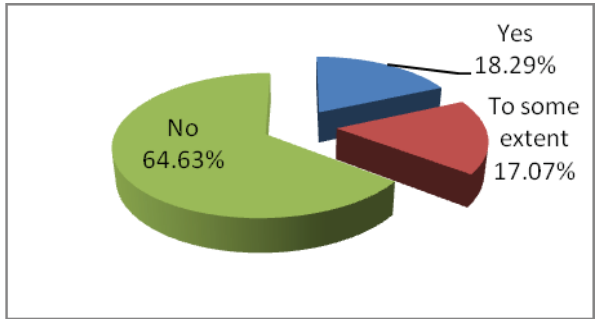 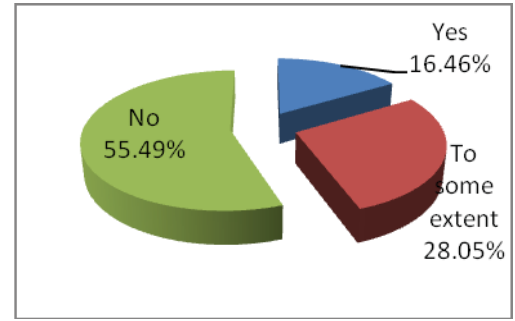

Figure 12. Difficulty Reading Texts: English.   Figure 13. Difficulty Reading Texts: Non-English.

## 4. CONCLUSIONS

A study was undertaken to determine the important usability factors (UF) related to the English and non-English language versions of an exemplar website. A questionnaire based on these fifteen UFs was used to assess the performance differences between the BBC news website and its non-English version.

The study found that in general, the English-language website gave a superior usability experience compared to its non-English version. In particular this was found to be true for all the usability factors except those related to the use of graphics. The overall conclusion of the study is that more effort needs to be spent in addressing the non-English websites so that their usability is at least equal to the English version. This is very important as some of the users are in fact bilingual and their negative impression of the non-English version can have an adverse image of the corporation producing the website. This may consequently harm the company's international reputation and market penetration.